\documentstyle[amssymb]{article}

\input tcilatex

\begin{document}

\author{Nicholas Ionescu-Pallas, Marius I. Piso and Silvia Sainar \\
Institute of Gravitation and Space Sciences, 71111 Bucharest}
\title{Solar System test for the existence of gravitational waves }
\date{ \ }
\maketitle

\begin{abstract}
Starting from a static spherically symmetric solution of the Einstein's
field equations in the second approximation in the perfect fluid scheme, in
the exterior of the source, the corresponding metrics depend on a
dimensionless parameter $\alpha $ . Taking the Sun for the source, the
influence of $\alpha $ on the classical Solar System tests of the general
relativity is estimated, the only $\alpha $-dependent one being the Shapiro
radar-echo delay experiment. Performing such a test in given conditions
within a precision better than $10^{-2}$, it is possible to obtain an
experimental value for $\alpha $ . If $\alpha =1$, the Einstein's equations
in the weak field approximation take the D'Alembert form, which attests the
existence of gravitational waves\QQfnmark{%
This work has been partially supported by the 1033C Romanian Space Agency
contract}.

\bigskip\ 

\end{abstract}

\section{Introduction}

For the metric theories of gravitation, the retardation of light signal
passing near massive bodies provides a principal test (Shapiro, 1964-1989,
Weinberg, 1972, Reasenberg et al., 1979). In the following, we shall perform
a more accurate analysis of this effect, starting from its ''anisotropy'',
i.e. the direction dependence of the light velocity in the gravitational
field of the Sun - which distinguishes it from other classical Solar System
tests. The origin of the anisotropy is to be found in the manner of gauging
the gravitational potentials, the effect being evidenced by means of a
certain dimensionless parameter $\alpha $. In the second section, a
statically symmetric solution of the Einstein's field equations in the
second approximation for a spherical mass distribution is given and the
equations of motion for both massive bodies and photons are derived. In the
third section, the influence of this metrics on the Solar System tests of
General Relativity is estimated. The fourth section concerns with the
connection between the Shapiro test and the propagation of gravitational
waves.

\section{The static field of the Sun and the equations of motion}

\subsection{The static metrics}

We start from the standard static spherically symmetric metrics (Zeldovich,
Novikov, 1971): 
\begin{equation}  \label{1}
(dS)^2=e^{\nu (r)}(cdt)^2-\left[ e^{\lambda (r)}(dr)^2+e^{\sigma
(r)}r^2d\Omega \right]
\end{equation}
where $\nu (r)$, $\lambda (r)$ and $\sigma (r)$ are arbitrary functions to
be established from physical arguments. By asking this diagonal metric to
fulfill the Einstein's field equations: 
\begin{equation}  \label{2}
E_{\mu v}\equiv R_{\mu \nu }-\frac 12g_{\mu \nu }R=-\frac{8\pi G}{c^4}T_{\mu
\nu }\quad \quad \mu ,\nu =0,1,2,3
\end{equation}
in the perfect fluid scheme (Misner, Thorne, Wheeler, 1973), we come to the
following expressions for the nonzero components of $E_{\mu \nu }$ and $%
T_{\mu \nu }$: 
\begin{equation}  \label{3}
\begin{array}{c}
E_{00}=e^{\nu -\lambda }\left[ (\Delta \sigma +\frac 34\sigma ^{\prime
2})+\frac 1r(\sigma ^{\prime }-\lambda ^{\prime })-\frac 12\lambda ^{\prime
}\sigma ^{\prime }+\frac 1{r^2}\right] -\frac 1{r^2}e^{\nu -\sigma } \\ 
\\ 
E_{11}=-\left[ \frac 14\sigma ^{\prime 2}+\frac 12\nu ^{\prime }\sigma
^{\prime }+\frac 1r(\sigma ^{\prime }+\nu ^{\prime })+\frac 1{r^2}\right]
+\frac 1{r^2}e^{\lambda -\sigma } \\ 
\\ 
E_{22}=r^2e^{\sigma -\lambda }\left[ -\frac 12(\Delta \sigma +\frac 12\sigma
^{\prime 2})-\frac 12(\Delta \nu +\frac 12\nu ^{\prime 2})+\frac 1{2r}(\nu
^{\prime }+\lambda ^{\prime })+\frac 14\sigma ^{\prime }(\lambda ^{\prime
}-\nu ^{\prime })+\frac 14\nu ^{\prime }\lambda ^{\prime }\right] \\ 
\\ 
E_{33}=(\sin {}^2\theta )\cdot E_{22} \\ 
\\ 
T_{00}=e^\nu \left[ \rho c^2+\left( \rho \int_0^p \frac{dp}\rho -p\right) %
\right] \\ 
\\ 
T_{11}=pe^\lambda \qquad T_{22}=r^2pe^\sigma \qquad T_{33}=(\sin {}^2\theta
)T_{22}%
\end{array}%
\QQfntext{0}{
This work has been partially supported by the 1033C Romanian Space Agency
contract}
\end{equation}
where $\rho $ is the invariant rest mass density and $p$ is the invariant
pressure of the source fluid; we denoted by prime the derivative with
respect to $r$ . In the linear approximation, the field equations become: 
\begin{equation}  \label{4}
\begin{array}{c}
\Delta \sigma +\frac 1r(\sigma ^{\prime }-\lambda ^{\prime })+\frac
1{r^2}(\sigma -\lambda )=- \frac{8\pi G}{c^2}\rho \\ 
\\ 
-\frac 1r(\sigma ^{\prime }+\nu ^{\prime })-\frac 1{r^2}(\sigma -\lambda )=0
\\ 
\\ 
\Delta \sigma +\Delta \nu -\frac 1r(\nu ^{\prime }+\lambda ^{\prime })=0%
\end{array}%
\end{equation}
Assuming the source to be a sphere of radius $R$ , we have: 
\begin{equation}  \label{5}
M_0=\frac 1{c^2}\int_0^R\rho _E(r)\cdot 4\pi r^2dr\qquad \Delta \Phi =-4\pi
\frac G{c^2}\rho _E
\end{equation}
where $\rho _E$ is the total rest energy density of the source, $M_0$ is the
total rest mass and $\Phi $ the Newtonian potential. In the linear
approximation, $\rho _E=\rho c^2$. In the second order approximation, the
gravitational self energy as well as the elastic energy coming from the
equilibrium pressure are included. The solution of \ref{4} is: 
\begin{equation}  \label{6}
\nu =-2\frac \Phi {c^2}\qquad \lambda =(r\sigma )^{\prime }-2\frac{r\Phi
^{\prime }}{c^2}\qquad \sigma -\text{arbitrary}
\end{equation}
(see Tonnelat, 1965 about the arbitrariness of $\sigma $).

Outside the mass distribution, an accurate solution of the field equations $%
E_{\mu \nu }=0$ is: 
\begin{equation}  \label{7}
(dS)^2=\left[ 1-2\frac \mu {f(r)}\right] (cdt)^2-\left\{ \left[ 1-2\frac \mu
{f(r)}\right] ^{-1}f^{\prime 2}(r)(dr)^2+f^2(r)d\Omega \right\}
\end{equation}
where, due to the asymptotic flatness, $f(r)$ is subjected to the
constraint: 
\begin{equation}  \label{8}
\lim _{r\rightarrow \infty }\frac{f(r)}r=1
\end{equation}
The solution \ref{7} may be checked by direct standard calculations. The
integration constant $\mu $ is determined as: 
\begin{equation}  \label{9}
\mu =\frac{GM_0}{c^2}\quad \text{,}
\end{equation}
from the necessity of regaining Newtonian theory as a non-relativistic
limit. From \ref{8} we can write $f(r)$ as an expansion: 
\begin{equation}  \label{10}
f(r)=r+\alpha \mu +O(\frac{\mu ^2}r)
\end{equation}
provided that $r$ and $\mu $ are the only arguments of $f$ with physical
dimensions. Here $\alpha $ is a dimensionless parameter. Using \ref{10} we
get an explicit approximate form of the metrics \ref{7}: 
\begin{equation}  \label{11}
(dS)^2=\left[ 1-2\frac \mu r+2\alpha \left( \frac \mu r\right) ^2\right]
(cdt)^2-\left[ \left( 1+2\frac \mu r\right) (dr)^2+\left( 1+2\alpha \frac
\mu r\right) r^2d\Omega \right]
\end{equation}

A little discussion is here needed. Outside the source, derived from \ref{1}
and \ref{11}, we have $\sigma \approx 2\alpha \frac \mu r$. On the other
side, the quantities $\nu $, $\lambda $, $\sigma $ cannot contain $\Phi
^{\prime \prime }$, due to difficulties concerning the continuity conditions
on the surface of the spherical source. The only possibility in hand is
therefore $\sigma \approx 2\alpha \Phi $ , which leads to: 
\begin{equation}  \label{12}
\begin{array}{c}
(dS)^2\approx \left( 1-2\frac \Phi {c^2}+2\alpha \frac{\Phi ^2}{c^4}\right)
(cdt)^2- \\ 
\\ 
-\left\{ \left[ 1+2\alpha \frac \Phi {c^2}-2(1-\alpha )\frac{r\Phi ^{\prime
} }{c^2}\right] (dr)^2+\left( 1+2\alpha \frac \Phi {c^2}\right) r^2d\Omega
\right\}%
\end{array}%
\end{equation}
The going through the solution \ref{12} ensures the compatibility between
the inner and outer solutions within the given approximation. Actually, it
was proved that the metrics \ref{12} is a second order approximate solution
of the field equations \ref{2} provided that the rest energy density and the
equilibrium pressure are: 
\begin{equation}  \label{13}
\rho _E=\left( 1-\frac{GM_0}{c^2R}\right) \rho c^2+(\alpha -1)(\Phi +2r\Phi
^{\prime })\rho +2p\quad \quad p=-\int_r^R\rho \Phi ^{\prime }dr
\end{equation}
Details are given in Appendix A.

\subsection{The equations of motion}

The equations of motion are resulting from the geodetic variational
principle, which is equivalent to the Euler-Lagrange variational principle: 
\begin{equation}  \label{14}
\delta \int {\em L}dt=0\qquad {\em L}\equiv -m_0c^2{\em K}\qquad {\em K}%
\equiv \frac{dS}{cdt}=\left[ A-\left( B\frac{{\bf v}^2}{c^2}+C\frac{({\bf v}%
\cdot {\bf r})^2}{c^2r^2}\right) \right] ^{1/2}
\end{equation}
\begin{equation}  \label{15}
\begin{array}{c}
\frac d{dt}\left( \frac{\partial {\em L}}{\partial {\bf v}}\right) -\frac{%
\partial {\em L}}{\partial {\bf r}}=0\qquad \frac{d{\bf p}}{dt}-{\bf F}=0 \\ 
\\ 
{\bf p}=\frac{\partial {\em L}}{\partial {\bf v}}=-\frac 12E\frac 1A\frac
\partial {\partial {\bf v}}({\em K}^2)\qquad {\bf F}=\frac{\partial {\em L}}{%
\partial {\bf r}}=-\frac 12E\frac 1A\frac \partial {\partial {\bf r}}({\em K}%
^2)%
\end{array}%
\end{equation}
Here we used the energy integral: 
\begin{equation}  \label{16}
{\em E}={\bf v}\cdot {\bf p}-{\em L}=\frac{m_0c^2}{{\em K}}A
\end{equation}
in order to eliminate the quantity $m_0{\em K}^{-1}$ in the expressions of
the momentum ${\bf p}$ and the force ${\bf F}$ , because $m_0{\em K}^{-1}$
becomes undetermined, $0/0$, in the case of the photon. From \ref{14} and %
\ref{15}, we get: 
\begin{equation}  \label{17}
{\bf p}=-\frac 12\frac{m_0c^2}{{\em K}}\frac \partial {\partial {\bf v}}{\em %
K}^2\qquad {\bf F}=-\frac 12\frac{m_0c^2}{{\em K}}\frac \partial {\partial 
{\bf r}}({\em K}^2)
\end{equation}
As $d{\bf E}/dt=0$, the equation of motion acquires a form which holds for
both massive bodies and photons: 
\begin{equation}  \label{18}
\frac d{dt}\left[ \frac 1A\frac \partial {\partial {\bf v}}\left( \frac 12%
{\em K}^2\right) \right] -\frac 1A\frac \partial {\partial {\bf r}}\left(
\frac 12{\em K}^2\right) =0
\end{equation}
Comparing $dS/cdt$ from \ref{14} to the homologous quantity from \ref{11},
we get: 
\begin{equation}  \label{19}
A\approx 1-2\frac \mu r+2\alpha \left( \frac \mu r\right) ^2\qquad B\approx
1+2\alpha \frac \mu r\qquad C\approx -2(\alpha -1)\frac \mu r
\end{equation}
By using series expansions relying on $\mu /r\ll 1$, we finally derived the
acceleration experienced by a point-like body of rest mass $m_0$, $m_0\ll
M_0 $ , outside the source: 
\begin{equation}  \label{20}
{\bf a}\approx -\frac{GM_0}{r^3}\left[ 1-2C_1\frac \mu r+2C_2{\bf \beta }%
^2-2C_3\frac{({\bf \beta }\cdot {\bf r})^2}{r^2}\right] {\bf r}+2C_4 \frac{%
GM_0}{r^3}({\bf \beta }\cdot {\bf r}){\bf \beta }
\end{equation}
where: 
\begin{equation}  \label{21}
C_1=\alpha +1\qquad C_2=1-\frac \alpha 2\qquad C_3=-\frac 32(\alpha
-1)\qquad C_4=\alpha +1\qquad {\bf \beta }=\frac{{\bf v}}c
\end{equation}

\section{The Solar System general relativistic tests}

From the results of section II, performing usual computations (Weinberg,
1972), we derived formulas for estimating the general relativistic tests,
within the Solar System, for an arbitrary gauge condition. The results are
shown in the sequel.

The {\em perihelion advance} of Mercury per revolution $\delta \omega $ may
be estimated by a standard procedure. The equation \ref{20} is integrated in
order to obtain the energy and angular momentum integrals: 
\begin{equation}  \label{22}
\begin{array}{c}
\left( \frac 12 {\bf v}^2-\frac{GM_0}r\right) +\frac{GM_0}{c^2r}%
[(2C_4-2C_2+\frac 43C_3){\bf v}^2+\frac 23C_3\frac{({\bf v\cdot r})^2}{r^2}
\\ 
\\ 
+(C_1+2C_2-2C_3-2C_4)] \frac{GM_0}r={\cal E} \\ 
\\ 
\left( 1+2C_4 \frac{GM_0}{c^2r}\right) \left( {\bf v\times r}\right) ={\bf A}
\\ 
\\ 
{\cal E}=4\frac E{m_0}-\left( \frac 52c^2+\frac 32\frac{E^2}{m_0^2c^2}%
\right) \quad ,\quad \quad \quad {\bf A}=\frac{c^2}E{\bf M}\quad%
\end{array}%
\end{equation}
where $\frac 12{\bf A}$ is the constant areolar velocity, $E$ the energy and 
${\bf M}$ the angular momentum. Out of these equations, a relativistic Binet
equation may be derived (Ionescu-Pallas, 1980): 
\begin{equation}  \label{23}
\frac{d^2(1/r)}{d\theta ^2}+k\cdot \left( \frac 1r\right) =\frac 1{p_1}+2C_3%
{\em \epsilon }^2\frac \mu {p_0^2}\cos {}^2\theta
\end{equation}
where $k=1+2\cdot \frac \mu {p_0}\cdot (C_1-2C_2-2C_4)$, $\mu =GM_0/c^2$, $%
p_0=a(1-{\em \epsilon }^2)$, $p_1=p_0+O(\mu )$. Then, a perturbational
technique is used for reaching the trajectory formula: 
\begin{equation}  \label{24}
\frac 1r=\frac 1{kp_1}\left[ 1+{\em \epsilon }\cos \left( \theta \sqrt{k}%
\right) \right] +\frac 23C_3{\em \epsilon }^2\frac \mu {p_0^2}\left( 1+\sin
{}^2\theta \right) \quad .
\end{equation}
After such calculations, $\delta \omega $ turns out to have the expression: 
\begin{equation}  \label{25}
\delta \omega =(-C_1+2C_2+2C_4)\frac{2\pi GM_0}{c^2a(1-{\em \epsilon }^2)}=%
\frac{6\pi GM_0}{c^2a(1-{\em \epsilon }^2)}
\end{equation}
which is independent on $\alpha $. Here $M_0$ is the rest mass of the Sun, $%
a $ - semi major axis of the ecliptic of Mercury, ${\em \epsilon }$ -
Mercury ecliptic eccentricity.

The {\em light deflection} angle $\delta \Psi $ is defined as: 
\begin{equation}  \label{26}
\delta \Psi =-\frac 1c\int_{-\infty }^{+\infty }a_x(t)dt=-\frac 1c\left[
v_x(+\infty )-v_x(-\infty )\right]
\end{equation}
where: 
\begin{equation}  \label{27}
a_x=-\frac{GM_0}{r^3}R\left[ (1+2C_2)-2C_3\cos {}^2\theta \right] \quad
\quad r=\left( c^2t^2+R^2\right) ^{\frac 12}\quad \quad \cos \theta = \frac{%
ct}R
\end{equation}
Here, $R$ is the Sun radius; $a_x$ is obtained from \ref{20} by replacing
there ${\bf v\sim }c{\bf j}$ and $x=R$, $y=ct$. $v_x$ is the projection of
the photon velocity on the $Ox$ axis (transverse to the trajectory)
(Ionescu-Pallas, 1980). After standard calculations, we get: 
\begin{equation}  \label{28}
\delta \Psi =\left( 1+2C_2-\frac 23C_3\right) \frac{2GM_0}{c^2R}= \frac{4GM_0%
}{c^2R}\text{\quad ,}
\end{equation}
that is $\delta \Psi $ is also independent on $\alpha $ .

Obviously, the {\em gravitational red shift} is independent on $\alpha $ ,
due to its connection only to the linear approximation of $g_{00}$ .

However, the situation turns not to be the same with the {\em Shapiro
retardation effect} (see for the following Shapiro, 1964, 1989, Weinberg,
1972, Misner, Thorne, Wheeler, 1973). We shall derive the formula for $%
\delta T$ - the retardation of a radar signal in a to-and-fro Terra-Mercury
journey, when Mercury is at the upper conjunction. The elementary duration
spent by a photon is, from \ref{11}: 
\begin{equation}  \label{29}
cdt\approx \left[ 1+2\frac \mu r+(\alpha -1)R^2\frac \mu {r^3}\right] \frac{%
rdr}{\sqrt{r^2-R^2}}+R\frac{d(\theta -\theta _0)}{dr}dr
\end{equation}
The photon is travelling, let's say, almost along the $Oy$ axis, passing at
the nearest distance $R$ from the center of the Sun. The first term in \ref%
{29} leads to the retardation due to the gravitational refractive index $n$
along the non-deviated trajectory ($x=R$) 
\begin{equation}  \label{30}
n=1+2\frac \mu r+(\alpha -1)R^2\frac \mu {r^3}
\end{equation}
The second term leads to the ''overtaking'' retardation. There are no
problems to estimate the first effect. For the estimation of the second
effect, we first notice that $\theta _0(r)=\arccos \frac Rr$ is the equation
of the line $x=R$ (the trajectory in the absence of the gravitational
field). Thereafter, we make use of the trajectory equation: 
\begin{equation}  \label{31}
\frac 1r\approx \frac{\cos \theta }R+\frac \mu {R^2}\left[ 2-(\alpha +1)\cos
\theta +(\alpha -1)\cos {}^2\theta \right]
\end{equation}
in order to get the formula: 
\begin{equation}  \label{32}
(\theta -\theta _0)\approx \frac \mu R\frac r{\sqrt{r^2-R^2}}\left[
2-(\alpha +1)\frac Rr+(\alpha -1)\left( \frac Rr\right) ^2\right]
\end{equation}
whence 
\begin{equation}  \label{33}
R\frac d{dr}(\theta -\theta _0)=\frac{\mu R}{\sqrt{r^2-R^2}}\left[ \frac{%
\alpha +1}{r+R}-(\alpha -1)\frac R{r^2}\right]
\end{equation}
The integration is straightforward. We obtain for the retardation $\delta T$
the formula: 
\begin{equation}  \label{34}
\delta T=\frac{4GM_0}{c^3}\left[ \ln \left( \frac{4l_1l_2}{R^2}e^{\alpha
-1}\right) +2\right] =\left[ 220.5+19.7(\alpha +1)\right] \times 10^{-6}s
\end{equation}
where $l_1$ is the major semi-axis of the ecliptic of Terra and $l_2$ is the
major semi-axis of the ecliptic of Mercury. Performing a Shapiro-type
experiment within an error of $\leq 10^{-2}$ it is possible to determine an
experimental value of $\alpha $.

\section{The gauge parameter $\protect\alpha $ and the gravitational waves
propagation}

We write the general metric tensor as: 
\begin{equation}  \label{35}
g_{\mu \nu }=\eta _{\mu \nu }+h_{\mu \nu }
\end{equation}
where $\eta _{\mu \nu }=2\delta _{0\mu }\delta _{0\nu }-\delta _{\mu \nu }$
is the covariant Minkowski tensor. From \ref{11} and \ref{19}, we have, in
Cartesian coordinates: 
\begin{equation}  \label{36}
h_{00}=(A-1)\qquad h_{jk}=-(B-1)\delta _{jk}-C\frac{x_jx_k}{r^2}\quad
,\qquad j,k=1,2,3
\end{equation}
Imposing the well-known Hilbert gauge condition (Zeldovich, Novikov, 1971,
Tonnelat, 1965): 
\begin{equation}  \label{37}
\left( h^{\lambda \sigma }-\frac 12\eta ^{\lambda \sigma }h\right) ,_\lambda
=0\quad \quad \lambda ,\sigma =0,1,2,3\quad \quad (\quad )_{,\lambda }\equiv
\frac \partial {\partial x^\lambda }(\quad )
\end{equation}
\quad where $\eta ^{\lambda \sigma }=2\delta ^{0\lambda }\delta ^{0\sigma
}-\delta ^{\lambda \sigma }$ is the contravariant Minkowski tensor, with: 
\begin{equation}  \label{38}
h=\eta ^{\lambda \sigma }h_{\lambda \sigma }\approx 4\alpha \frac \mu r
\end{equation}
we get from \ref{36}, \ref{37} and \ref{38}, by replacing the coefficients $%
A,B,C$ with their values from \ref{19}: 
\begin{equation}  \label{39}
\alpha =1
\end{equation}
i.e. the condition for $\alpha $ in order to satisfy the Hilbert gauge.

From the other point of view, Einstein's field equations \ref{2}, written in
rectangular coordinates and in an inertia frame: 
\begin{equation}  \label{40}
R_{\mu v}=-\frac{8\pi G}{c^4}\left( T_{\mu \nu }-\frac 12g_{\mu \nu }T\right)
\end{equation}
reduce, with $T_{\mu \nu }$ in the perfect fluid scheme, and within the
linear approximation \ref{35} ($\left| h_{\mu \nu }\right| \ll 1$), to: 
\begin{equation}  \label{41}
\Box h_{\mu \nu }-\left[ \left( h_\mu ^\lambda -\frac 12\delta _\mu ^\lambda
h\right) ,_{\lambda \nu }+\left( h_\nu ^\lambda -\frac 12\delta _\nu
^\lambda h\right) ,_{\lambda \mu }\right] =-\frac{8\pi G}{c^2}\rho \left(
u_\mu u_\nu -\frac 12\eta _{\mu \nu }\right)
\end{equation}
where: 
\begin{equation}  \label{42}
\begin{array}{c}
u_\mu =\eta _{\mu \lambda }u^\lambda \qquad u^\lambda = \frac{dx^\lambda }{%
dS_M}\quad \quad (dS_M)^2=\eta _{\lambda \sigma }dx^\lambda dx^\sigma \\ 
\\ 
\Box \equiv \eta ^{\lambda \sigma }\frac \partial {\partial x^\lambda }\frac
\partial {\partial x^\sigma }=-\left( \triangle -\frac 1{c^2}\frac{\partial
^2}{\partial t^2}\right)%
\end{array}%
\end{equation}
For bringing \ref{40} to the D'Alembert form, we need just the condition \ref%
{37}, i.e. $\alpha =1$. The D'Alembert form unambiguously attests the
existence of gravitational waves, in the framework of Einstein's General
Relativity Theory.

\section{Conclusions}

We performed a more accurate analysis of the Shapiro time-delay Solar System
test of general relativity. In order to perform the analysis, in the second
section of the paper we studied a general spherically symmetric static
metrics given by the Sun, in the second approximation. Starting from a more
general solution of the Einstein's equations \ref{7} and imposing the
physical conditions of regaining the Newtonian limit and the continuity of
the solution on the surface of the spherical mass source, we obtained an
approximate form of the metrics in the exterior of the mass source \ref{11},
which contains, in the second approximation an undetermined dimensionless
parameter $\alpha $. By means of standard calculations performed using the
metrics \ref{11}, we got a form of the equations of motion \ref{18} for
point masses, which holds for both massive bodies and photons, and we
derived the acceleration \ref{20} experienced by a point-like massive body
or photon, both \ref{18} and \ref{20} holding for the movement in the static
field of the Sun. In the third section, we estimated the influence of the
undetermined gauge parameter $\alpha $ on the classical Solar System tests.
The perihelion advance, the light deflection and the gravitational red shift
are independent on $\alpha $. However, the Shapiro effect depends on $\alpha 
$. Performing the calculations for the retardation of the radar signal in a
to-and-fro Terra-Mercury journey, with Mercury at the upper conjunction, we
estimated a relative influence of $\alpha $ of the order $\sim 10^{-2}$ on
the total signal retardation of time \ref{34}. Finally, in the fourth
section, we established the connection of the parameter $\alpha $ to the
propagation of the gravitational waves. From the Hilbert gauge condition \ref%
{37}, necessary for the Einstein's equations in order to take the D'Alembert
form (which attests the existence of gravitational waves), we obtained the
value $\alpha =1$.

The short conclusions of our analysis are the following: {\em i)} it is
possible to determine an experimental value for $\alpha $ by means of a
Shapiro-type experiment, and {\em ii)} if $\alpha =1$, the existence of
general relativistic gravitational waves is proved.

Finally, a definite alternative - which may be solved experimentally through
the agency of the Shapiro effect - is reached: either the light propagation
in the gravitational field of the Sun is isotropic and the gravitational
waves do exist, or the respective propagation is anisotropic and the
gravitational waves are precluded. It is to point out that our conclusions
are conditioned by the hypothesis about the correctness of Einstein's
General Relativity Theory.

\ 

\begin{center}
{\LARGE REFERENCES}
\end{center}

\begin{quote}
Ionescu-Pallas, N.: 1980, {\em General Relativity and Cosmology} (in
Romanian), Ed. St. Enc., Bucharest.

Ionescu-Pallas, N., Piso, M.I., Sainar, S.: 1993, in {\em Proceedings of the
8th Int. Conf. of the Russian Gravitational Association}, RGA, Moscow, 47.

Misner, C.W., Thorne, K.S., Wheeler, J.A.: 1973, {\em Gravitation}, W.H.
Freeman \& Co..

Reasenberg, R.D. et al.: 1979, {\em Ap. J. Lett. 234, L219.}

Shapiro, I.I.: 1964, {\em Phys. Rev. Lett. 13, 789} (1964).

Shapiro, I.I. et al.: 1977, {\em J. Geophys. Res.} {\bf 82}, 4329.

Shapiro, I.I.: 1989, in {\em Proceedings of the 12th International
Conference on General Relativity and Gravitation}, Boulder.

Tonnelat, M.A.: 1965, {\em Les th\'eories unitaires de
l'\'el\'ectromagnetisme et de la gravitation}, Gauthier-Villars Ed..

Weinberg, S: 1972, {\em Gravitation and Cosmology}, Wiley.

Zeldovich, Ya.B., Novikov, I.D.: 1971, {\em Stars and Relativity}, vol. 1,
Chicago Univ. Press.

\newpage\ 
\end{quote}

\section{Appendix}

All the physical quantities entering the Einstein's field equations of
General Relativity Theory are expanded in power series of the constant $%
1/c^2 $: 
\begin{equation}  \label{a1a}
\nu =\frac 1{c^2}\nu _0+\frac 1{c^4}\nu _1+O\left( \frac 1{c^6}\right)
\end{equation}
\begin{equation}  \label{a1b}
\lambda =\frac 1{c^2}\lambda _0+\frac 1{c^4}\lambda _1+O\left( \frac
1{c^6}\right)
\end{equation}
\begin{equation}  \label{a1c}
\sigma =\frac 1{c^2}\sigma _0+\frac 1{c^4}\sigma _1+O\left( \frac
1{c^6}\right)
\end{equation}
\begin{equation}  \label{a1d}
\rho =\rho _M-\frac 1{c^2}\rho _M\left( \frac 12\lambda _0+\sigma _0\right)
+O\left( \frac 1{c^4}\right)
\end{equation}
\begin{equation}  \label{a1e}
p=p_M+O\left( \frac 1{c^2}\right)
\end{equation}
(the label $M$ stands for the Minkowskian nature of the respective physical
quantity). Accordingly, the tensors $E_{\alpha \beta }$ and $T_{\alpha \beta
}$ are themselves expanded in similar power series: 
\begin{equation}  \label{a2a}
E_{\alpha \beta }=\frac 1{c^2}E_{\alpha \beta }^{(0)}+\frac 1{c^4}E_{\alpha
\beta }^{(1)}+O(\frac 1{c^6})
\end{equation}
\begin{equation}  \label{a2b}
T_{\alpha \beta }=c^2T_{\alpha \beta }^{(0)}+T_{\alpha \beta }^{(1)}+O(\frac
1{c^2})
\end{equation}
and the field equations $E_{\alpha \beta }=-(8\pi G/c^4)T_{\alpha \beta }$
split in a set of equations which no more contain the constant $c$: 
\begin{equation}  \label{a3a}
E_{\alpha \beta }^{(0)}=-8\pi GT_{\alpha \beta }^{(0)}\quad ,
\end{equation}
\begin{equation}  \label{a3b}
E_{\alpha \beta }^{(1)}=-8\pi GT_{\alpha \beta }^{(1)}\quad \text{, etc...}
\end{equation}
One obtains: 
\begin{equation}  \label{a4a}
T_{\alpha \beta }^{(0)}=\rho _M(r)\delta _{0\alpha }\delta _{0\beta }\quad ,
\end{equation}
\begin{equation}  \label{a4b}
T_{\alpha \beta }^{(1)}=\frac 12\rho _M(r)\left[ \nu _0(R)+\nu _0(r)-\lambda
_0(r)-2\sigma _0(r)\right] \delta _{0\alpha }\delta _{0\beta }-a_{\alpha
\beta }p_M(r)
\end{equation}
where $a_{\alpha \beta }$ is the tensor of Minkowskian metric in spherical
coordinates, and $p_M(r)$ is the equilibrium pressure: 
\begin{equation}  \label{a5}
p_M(r)=-\int_r^R\rho _M(r)\Phi _M^{\prime }(r)dr
\end{equation}
expressed by means of the Newtonian potential: 
\begin{equation}  \label{a6}
\triangle _r\Phi _M=-4\pi G\rho _M
\end{equation}
The equations \ref{a3a}, written in explicit form, are the following:%
\[
\triangle _r\sigma _0+\frac 1r\left( \sigma _0^{\prime }-\lambda _0^{\prime
}\right) +\frac 1{r^2}\left( \sigma _0-\lambda _0\right) =-8\pi G\rho _M(r) 
\]
\begin{equation}  \label{a7}
-\frac 1r\left( \sigma _0^{\prime }+\nu _0^{\prime }\right) +\frac
1{r^2}\left( \lambda _0-\sigma _0\right) =0
\end{equation}
\[
-\frac 12\triangle _r\sigma _0-\frac 12\triangle _r\nu _0+\frac 1{2r}\left(
\nu _0^{\prime }+\lambda _0^{\prime }\right) =0 
\]

Assuming for $\sigma _0$ a value $\sigma _0=2\alpha \Phi _M$, one obtains
the solution of the system \ref{a7} as: 
\begin{equation}  \label{a8}
\nu _0=-2\Phi _M\quad ,\quad \quad \lambda _0=2\alpha \Phi _M(r)+2(\alpha
-1)r\Phi _M^{\prime }\quad ,\quad \quad \sigma _0=2\alpha \Phi _M
\end{equation}

The system \ref{a3b} contains not only the quantities $\nu _1,\lambda
_1,\sigma _1,$ but the quantities $\nu _0,\lambda _0,\sigma _0$ as well.
Inserting in \ref{a3b} the values \ref{a8} of the zero-labelled quantities,
one obtains \ref{a3b} in explicit form, namely: 
\begin{equation}  \label{a9a}
\begin{array}{c}
\left[ \triangle _r\sigma _1+\frac 1r\left( \sigma _1^{\prime }-\lambda
_1^{\prime }\right) +\frac 1{r^2}\left( \sigma _1-\lambda _1\right) \right]
+\left( \alpha ^2+2\alpha -2\right) \Phi _M^{\prime 2}= \\ 
\\ 
8\pi G\left\{ (\alpha -1)\left[ \Phi _M-(\alpha +1)r\Phi _M^{\prime }\right]
\rho _M+\left[ \frac{GM_0}R\rho _M+p_M\right] \right\}%
\end{array}%
\end{equation}
\begin{equation}  \label{a9b}
\left[ \frac 1{r^2}\left( \lambda _1-\sigma _1\right) -\frac 1r\left( \sigma
_1^{\prime }+\nu _1^{\prime }\right) \right] +\left( \alpha ^2-2\alpha
+2\right) \Phi _M^{\prime 2}=-8\pi G\rho _M\quad ,
\end{equation}
\begin{equation}  \label{a9c}
\begin{array}{c}
\left[ -\frac 12\triangle _r\sigma _1-\frac 12\triangle _r\nu _1+\frac
1{2r}\left( \nu _1^{\prime }+\lambda _1^{\prime }\right) \right] -\left(
\alpha ^2-2\alpha +2\right) \Phi _M^{\prime 2}= \\ 
\\ 
8\pi G\left\{ \frac 12\left( \alpha -1\right) ^2\rho _M\cdot r\Phi
_M^{\prime }-p_M\right\}%
\end{array}%
\end{equation}
We multiply \ref{a9c} by $2$ and thereafter add it with \ref{a9b} and \ref%
{a9a}; one obtains: 
\begin{equation}  \label{a10}
\triangle _r\nu _1+4\left( 1-\alpha \right) \Phi _M^{\prime 2}=-8\pi
G\left\{ \left( 1-\alpha \right) \left[ 2r\Phi _M^{\prime }-\Phi _M\right]
\rho _M+\left[ \frac{GM_0}R\rho _M-2p_M\right] \right\}
\end{equation}
Now, let us define a new function $\Phi $ by mean of the formula: 
\begin{equation}  \label{a11}
\Phi =\Phi _M+\frac 1{c^2}\left[ \left( \alpha -1\right) \Phi _M^2-\frac
12\nu _1\right]
\end{equation}
Accordingly, 
\begin{equation}  \label{a12}
1-2\frac \Phi {c^2}+2\alpha \frac{\Phi ^2}{c^4}=1+\frac 1{c^2}\nu _0+\frac
1{c^4}\left( \frac 12\nu _0^2+\nu _1\right) =e^\nu =g_{00}\quad .
\end{equation}
From \ref{a10} and \ref{a11} we derive the Poisson type equation for $\Phi $%
: 
\begin{equation}  \label{a13a}
\triangle _r\Phi =-4\pi \frac G{c^2}\rho _E\quad ,
\end{equation}
\begin{equation}  \label{a13b}
\rho _E=\left( 1-\frac{GM_0}{c^2R}\right) \rho _Mc^2+\left( \alpha -1\right)
\left( \Phi _M+2r\Phi _M^{\prime }\right) \rho _M+2p_M\quad .
\end{equation}

\end{document}